\documentstyle[psfig]{mn} 
\title{Photometry of neglected open clusters in the First and 
Fourth Galactic Quadrants}
\author[Carraro, Janes \& Eastman]        
{Giovanni Carraro$^{1,2}$, \thanks{Andes Fellow, on leave from Dipartimento 
di Astronomia, Universit\`a di Padova,
Vicolo Osservatorio 2, I-35122, Padova, Italy} Kenneth A. Janes$^3$ and
Jason D. Eastman$^{3,4}$
\thanks{email: 
gcarraro@das.uchile.cl (GC), janes@bu.edu (KAJ), jdeast@bu.edu (JDE)}\\ 
$^1$Departamento de Astronom\'ia, Universidad de Chile, 
Casilla 36-D, Santiago, Chile\\
$^2$Astronomy Department, Yale University, 
P.O. Box 208101, New Haven, CT 06520-8101 , USA\\
$^3$Department of Astronomy, Boston University, 725 Commonwealth Avenue, 
Boston, MA 02215, USA\\
$^4$Department of Astronomy, Ohio State University, 140 West 18th Avenue,
Columbus, OH 43210, USA\\
 } 
 
\date{\it Submitted: July 2005} 
\pubyear{2005} 
\begin{document} 
\maketitle 
\title{Open clusters in the First and Fourth Galactic Quadrant} 
 
\begin{abstract} 
CCD $BVI$ photometry is presented for 8 previously 
unstudied star clusters located
in the First and Fourth Galactic Quadrants:  AL~1, BH~150,
NGC~5764, Lynga~9, Czernik~37, BH~261, Berkeley~80 and King~25.
Color magnitude diagrams of the cluster regions suggest that several
of them (BH~150, Lynga~9, Czernik~37 and BH~261 and King~25) 
are so embedded in
the dense stellar population toward the 
galactic center that their properties, or even their existence as 
physical systems, cannot be confirmed.  Lynga~9, BH~261 and King~25 appear 
to be slight enhancements of dense star fields, BH~150 is probably just a 
single bright star in a dense field, and Czernik~37 may be a sparse, but
real cluster superimposed on the galactic bulge population.
We derive preliminary estimates of
the physical parameters for the remaining clusters. AL~1 appears to
be an intermediate age cluster beyond the solar circle on the far side
of the galaxy and the final two clusters, NGC 5764 and Berkeley 80 are
also of intermediate age but located inside the 
solar ring.  This set of clusters highlights the difficulties inherent in
studying the stellar populations toward the inner regions of the galaxy.

\end{abstract} 
 

\section{Introduction} 
According to the recent Galactic Open Clusters compilation 
by Dias et al. (2002, http://www.astro.iag.usp.br/~wilton),
1632 open clusters are known to exist in the Milky Way disk.
Unfortunately,  basic parameters like distance, reddening and age are
available for fewer than 
half the clusters in
this sample. This fact obviously limits the use of open clusters
as probes of the structure and  evolution of the Galactic disk.
A large observational effort is clearly needed to improve on
this situation.\\
This is particularly important for sparse, loose star clusters,
which are hard to distinguish from the rich Galactic field and may be
very close to the dissolution phase (Bonatto et al. 2004).\\
\noindent
The statistics of open cluster ages are dramatically skewed 
towards young star clusters, which are both more numerous and often
more visible (Wielen 1971).
However in recent years new efforts have been done
to provide observational material for intermediate-age and old star clusters
(Phelps et al. 1994, Kaluzny 1994, 
Hasegawa et al. 2004, Carraro et al 2005a, and references therein).
All this new observational material will surely result
in a revision of the open cluster age distribution 
and typical lifetime.\\
In this paper we present the first photometric study
of 8 overlooked, faint and highly contaminated 
open clusters located in the Fourth and First Galactic 
Quadrant having $305^o\leq l \leq 49^o$ and 
$-5^o.2\leq b \leq +5^o.9$ (see Table~1) and provide homogeneous derivation
of basic parameters using the Padova (Girardi et al. 2000)
family of isochrones.\\

\noindent
The plan of the paper is as follows. Sect.~2 describes
the observation strategy and reduction technique.
Sect.~3 deals with the Color-Magnitude Diagrams (CMD)
and illustrates the derivation of the clusters' fundamental
parameters. Finally, Sect.~4 provides a detailed discussion
of the results.

\begin{table}
\caption{Basic parameters of the clusters under investigation.
Coordinates are for J2000.0 equinox and have been 
visually re-determined by us.}
\begin{tabular}{ccccc}
\hline
\hline
\multicolumn{1}{c}{Name} &
\multicolumn{1}{c}{$RA$}  &
\multicolumn{1}{c}{$DEC$}  &
\multicolumn{1}{c}{$l$} &
\multicolumn{1}{c}{$b$} \\
\hline
& {\rm $hh:mm:ss$} & {\rm $^{o}$~:~$^{\prime}$~:~$^{\prime\prime}$} & [deg] & [deg]\\
\hline
AL 1          & 13:15:16 & -65:55:18 & 305.40 & -3.20\\
BH 150        & 13:38:04 & -63:20:45 & 308.11 & -0.93\\
NGC 5764      & 14:53:33 & -52:40:15 & 320.97 & +5.86\\
Lynga 9       & 16:20:41 & -48:32:00 & 334.51 & +1.08\\
Czernik 37    & 17:53:14 & -27:22:00 &   2.22 & -0.63\\
BH 261        & 18:14:05 & -28:38:00 &   3.35 & -5.29\\
Berkeley~80   & 18:54:21 & -01:13:00 &  32.17 & -1.25\\
King 25       & 19:24:35 & +13:42:14 &  48.86 & -0.93\\
\hline\hline
\end{tabular}
\end{table}

\begin{figure} 
\centerline{\psfig{file=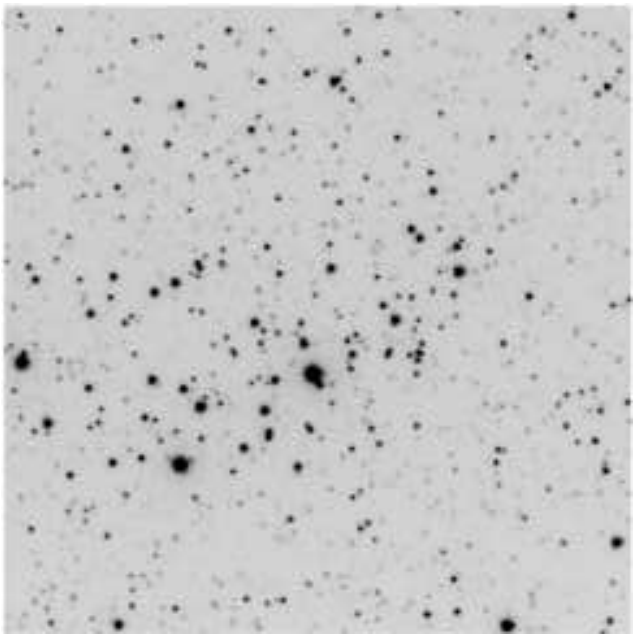,width=\columnwidth,angle=90}} 
\caption{I = 1200 sec image of A-L~1. North is up, East on the left,
and the covered area is $4^{\prime}.1 \times 4^{\prime}.1$.}
\end{figure}

\begin{figure} 
\centerline{\psfig{file=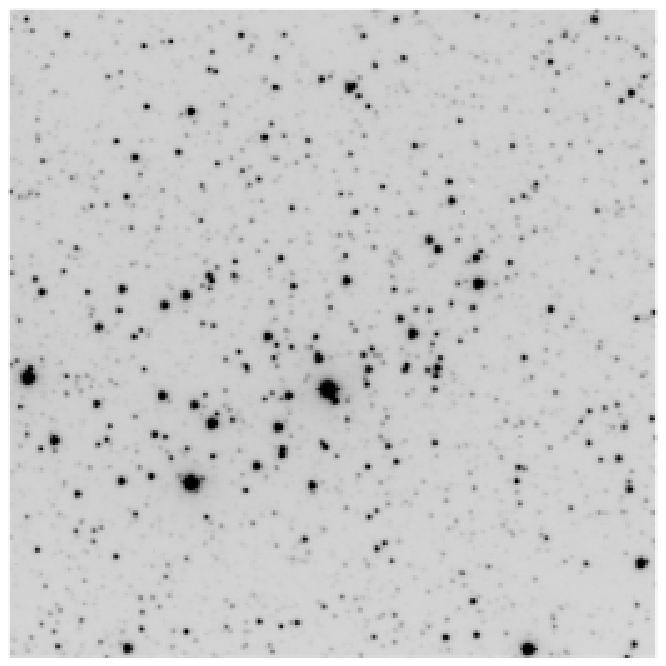,width=\columnwidth,angle=90}} 
\caption{I = 600 sec image of Berkeley~80. North is up, East on the left,
and the covered area is $4^{\prime}.1 \times 4^{\prime}.1$.}
\end{figure}

\begin{figure} 
\centerline{\psfig{file=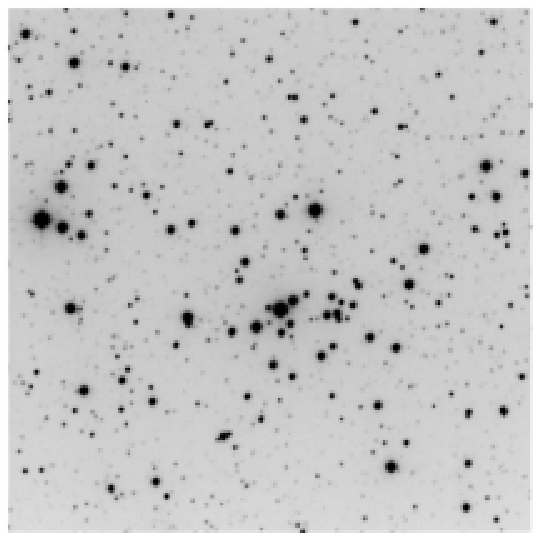,width=\columnwidth,angle=90}} 
\caption{I = 600 sec image of NGC~5764. North is up, East on the left,
and the covered area is $4^{\prime}.1 \times 4^{\prime}.1$.}
\end{figure}

\begin{figure} 
\centerline{\psfig{file=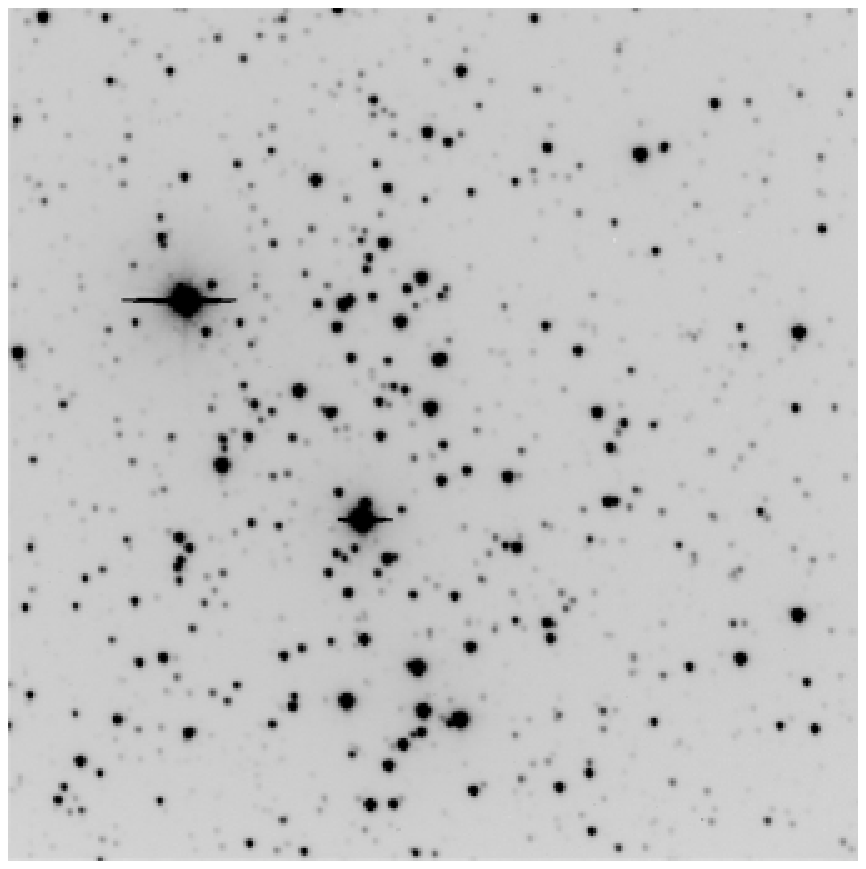,width=\columnwidth,angle=90}} 
\caption{I = 600 sec image of Czernik~37. North is up, East on the left,
and the covered area is $4^{\prime}.1 \times 4^{\prime}.1$.}
\end{figure} 

\begin{figure} 
\centerline{\psfig{file=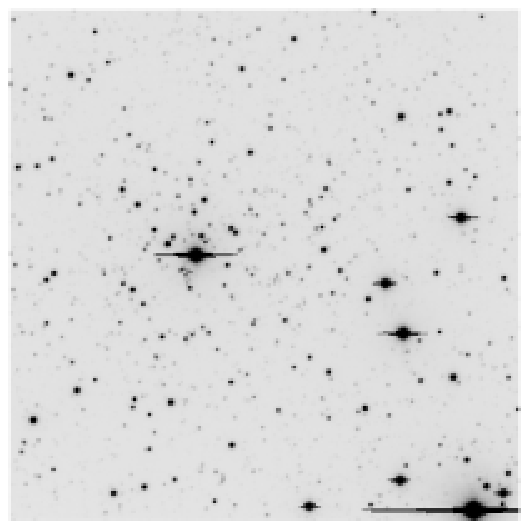,width=\columnwidth,angle=90}} 
\caption{I = 600 sec image of BH~150. North is up, East on the left,
and the covered area is $4^{\prime}.1 \times 4^{\prime}.1$.}
\end{figure} 

\begin{figure} 
\centerline{\psfig{file=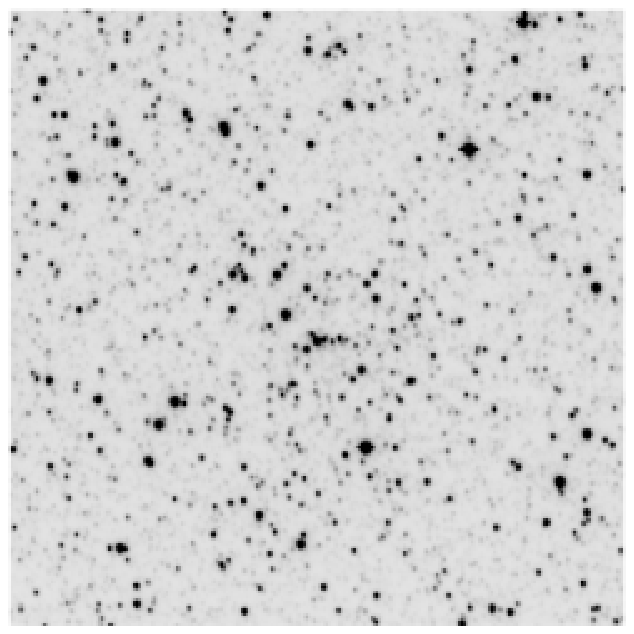,width=\columnwidth,angle=90}} 
\caption{I = 600 sec image of BH~261. North is up, East on the left,
and the covered area is $4^{\prime}.1 \times 4^{\prime}.1$.}
\end{figure}

\begin{figure} 
\centerline{\psfig{file=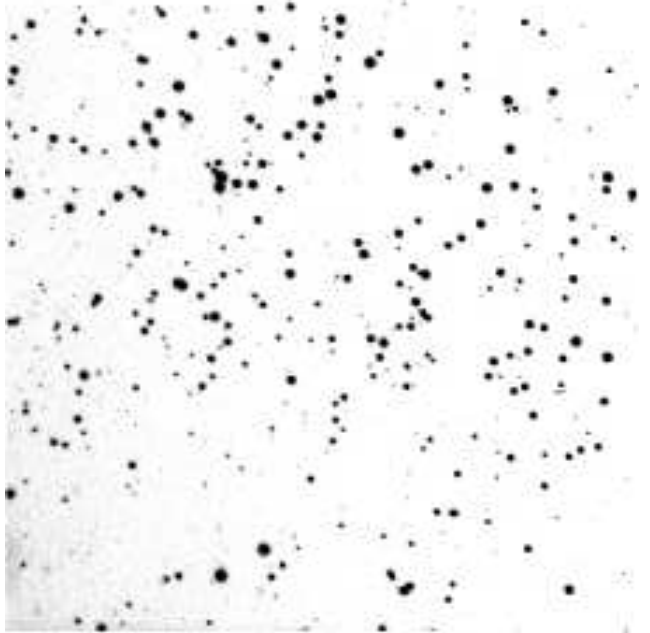,width=\columnwidth,angle=90}} 
\caption{B = 900 sec image of Lynga~9. North is up, East on the left,
and the covered area is $4^{\prime}.1 \times 4^{\prime}.1$.}
\end{figure}

\begin{figure} 
\centerline{\psfig{file=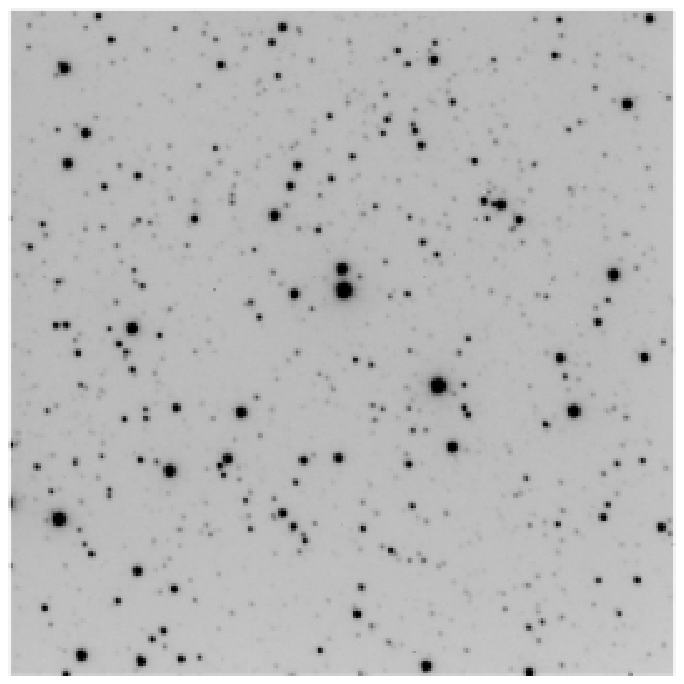,width=\columnwidth,angle=90}} 
\caption{I = 600 sec image of King~25. North is up, East on the left,
and the covered area is $4^{\prime}.1 \times 4^{\prime}.1$.}
\end{figure}

\section{Observations and Data Reduction} 
 
$\hspace{0.5cm}$
CCD $BVI$ observations were carried out with the CCD camera on-board
the  1. 0m telescope at Cerro Tololo Interamerican Observatory (CTIO,Chile), on
the night of
June 6, 2005.
With a pixel size of $0^{\prime\prime}.469$,  and a CCD size of 512 $\times$ 512
pixels,
this samples a $4^\prime.1\times4^\prime.1$ field on the sky.\\
\noindent
The details of the observations are listed in Table~2 where the observed
fields are
reported together with the exposure times, the average seeing values and the
range of air-masses during the observations.
Figs.~1 to 8 show deep I images in the area of
the clusters we observed.

\noindent
The data have been reduced with the
IRAF\footnote{IRAF is distributed by NOAO, which are operated by AURA under
cooperative agreement with the NSF.}
packages CCDRED, DAOPHOT, ALLSTAR and PHOTCAL using the point spread function 
(PSF) method (Stetson 1987).
The  night turned out to be photometric and very stable.
We derived calibration equations for all the 80 standard stars
observed during the night in the Landolt
(1992)  fields PG~1047+003, MarkA, PG~1323-085,
PG~1633+099, PG~1657+078 and  PG~2213-006 (see Table~2 for details).
Together with the clusters, 
we observed two control fields 20 arcmins
South of King~25, at 19:24:34.9, +13:22:15.3 (J2000.0), and 10 arcmins
North of Lynga~9,
at 16:20:40.8, -48:21:45.1 (J2000.0), to deal with field star
contamination. 
In fact these are the only two clusters which extend
beyond the field covered by the CCD.
Exposure of 600 secs in V and I were secured for these
fields.

\begin{table}
\fontsize{8} {10pt}\selectfont
\tabcolsep 0.10truecm
\caption{Journal of observations of clusters
and standard star fields (June 6, 2005).}
\begin{tabular}{cccccc}
\hline
\multicolumn{1}{c}{Field}         &
\multicolumn{1}{c}{Filter}        &
\multicolumn{1}{c}{Exposure time} &
\multicolumn{1}{c}{Seeing}        &
\multicolumn{1}{c}{Airmass}       \\
 & & [sec.] & [$\prime\prime$] & \\
\hline
AL~1       & B &       120,1200,1800   &   1.0 & 1.24 \\
            & V &        30,600,1200    &   1.0 & 1.24 \\
            & I &        30,600,1200    &   1.0 & 1.25 \\
\hline
BH~150      & B &      120,1200   &   0.9 & 1.20 \\
            & V &       30,600    &   0.8 & 1.20 \\
            & I &       30,600    &   0.8 & 1.21 \\
\hline
NGC~5764     & B &      120,1200  &   0.9 & 1.09 \\
             & V &      30,600    &   1.0 & 1.09 \\
             & I &      30,600    &   1.0 & 1.10 \\
\hline
Lynga~9      & B &    120,900    &   1.0 & 1.07 \\
             & V &      30,600   &   1.0 & 1.06 \\
             & I &      30,600   &   0.8 & 1.05 \\
\hline
Czernik~37    & B &    120,900    &   1.1 & 1.00 \\
              & V &      30,600   &   1.0 & 1.00 \\
              & I &      30,600   &   1.0 & 1.01 \\
\hline
BH~261        & B &    120,900    &   1.0 & 1.01 \\
              & V &      30,600   &   0.9 & 1.01 \\
              & I &      30,600   &   0.9 & 1.02 \\
\hline
Berkeley~80     & B &    120,900   &   0.8 & 1.16 \\
                & V &      30,600  &   0.8 & 1.19 \\
                & I &      30,600  &   0.8 & 1.21 \\
\hline
King~25       & B &    120,1200    &   1.0 & 1.43 \\
              & V &      30,600    &   1.0 & 1.49 \\
              & I &      30,600    &   1.0 & 1.52 \\
\hline
MarkA         & B &   $2 \times$120   &   1.0 & 1.24-1.46 \\
              & V &   $2 \times$40    &   1.0 & 1.24-1.46 \\
              & I &   $2 \times$20    &   1.1 & 1.24-1.46 \\
\hline
PG 1323-085   & B &   $3 \times$120   &   1.1 & 1.00-1.91 \\
              & V &   $3 \times$40    &   1.1 & 1.00-1.91 \\
              & I &   $3 \times$20    &   1.0 & 1.00-1.91 \\
\hline
PG 1633+099   & B &   $3 \times$120   &   1.1 & 1.20-1.74 \\
              & V &   $3 \times$40    &   1.1 & 1.20-1.74 \\
              & I &   $3 \times$20    &   1.0 & 1.20-1.74 \\
\hline
PG 1657+078   & B &   $4 \times$120   &   1.1 & 1.06-1.34 \\
              & V &   $4 \times$40    &   1.0 & 1.06-1.34 \\
              & I &   $4 \times$20    &   0.9 & 1.06-1.34 \\
\hline
PG 2213-006   & B &   $3 \times$120   &   1.1 & 1.00-1.54 \\
              & V &   $3 \times$40    &   1.0 & 1.00-1.54 \\
              & I &   $3 \times$20    &   1.0 & 1.00-1.54 \\
\hline
PG 1047+003   & B &   $3 \times$120   &   1.1 & 1.18-1.47 \\
              & V &   $3 \times$40    &   0.9 & 1.18-1.47 \\
              & I &   $3 \times$20    &   0.9 & 1.18-1.47 \\
\hline
\hline
\end{tabular}
\end{table}

\noindent
The calibration equations are of the form:\\

\noindent
$ b = B + b_1 + b_2 \times X + b_3~(B-V)$ \\
$ v = V + v_1 + v_2 \times X + v_3~(B-V)$ \\
$ v = V + v_{1,i} + v_{2,i} \times X + v_{3,i} \times (V-I)$ \\
$ i = I + i_1 + i_2 \times X + i_3~(V-I)$ ,\\

\noindent

\begin{table}
\tabcolsep 0.2truecm
\caption {Coefficients of the calibration equations}
\begin{tabular}{ccc}
\hline

$b_1 = 3.573 \pm 0.009$ & $b_2 =  0.25 \pm 0.02$ & $b_3 = -0.155 \pm 0.008$ \\
$v_1 = 3.447 \pm 0.005$ & $v_2 =  0.16 \pm 0.02$ & $v_3 = -0.019 \pm 0.005$ \\
$v_{1,i} = 3.448 \pm 0.005$ & $v_{2,i} =  0.16 \pm 0.02$ & $v_{3,i} =  -0.016 \pm0.005$ \\
$i_1 = 4.338 \pm 0.005$ & $i_2 =  0.08 \pm 0.02$ & $i_3 =  -0.022 \pm 0.005$ \\

\hline
\end{tabular}
\end{table}

\noindent
where $BVI$ are standard magnitudes, $bvi$ are the instrumental ones and  $X$ is
the airmass; all the coefficient values are reported in Table~3.
The standard
stars in these fields provide a very good color coverage.
The final global {\it r.m.s.} (calibration plus DAOPHOT internal errors)
are 0.033, 0.031 and 0.031 for the B, V and I filters,
respectively (Patat \&  Carraro 2001).

\noindent
We generally used the third equation to calibrate the $V$ magnitude
in order to get the same magnitude depth both in the cluster
and in the field.
The limiting magnitudes are B = 21.9, V = 22.5
and I =21.8.
Moreover we performed a completeness analysis following the method described in
Baume et al. (2005). It turns out that our sample
has completeness level larger than 50$\%$ down to B = 20.0, V = 21.0
and I = 20.5.

The final photometric catalogs for
(coordinates,
B, V and I magnitudes and errors)
consist of 3392, 1949, 1537, 1373, 1178, 3729, 1738 and 883 stars
for AL~1, BH~150, NGC~5764, Lynga~9, Czernik~37, BH~261, Berkeley~80 and 
King~25,
respectively, and are made
available in electronic form at the
WEBDA\footnote{http://obswww.unige.ch/webda/navigation.html} site
maintained by J.-C. Mermilliod.\\

     \begin{figure*}
     \centerline{\psfig{file=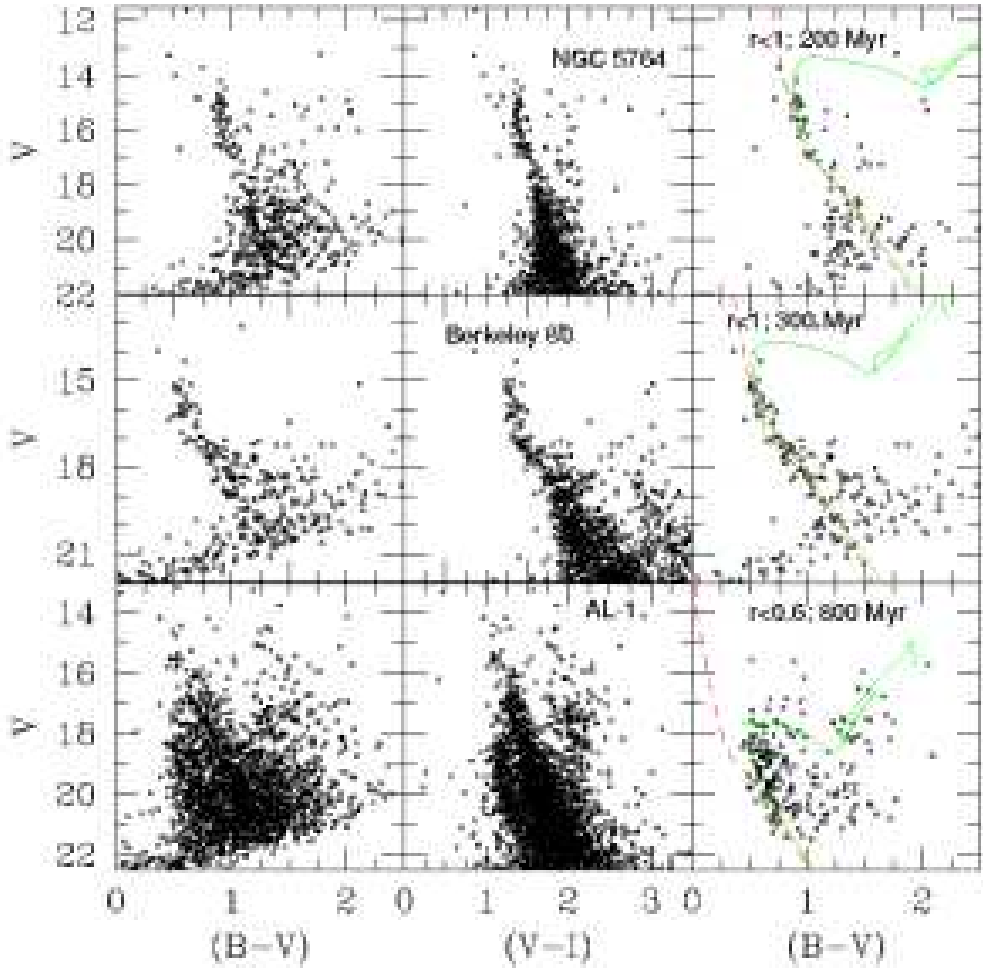,width=14cm}}
     \caption{
     CMDs of the stars in the field of AL~1, Berkeley~80 and NGC~5764. 
     {\bf Left panels}: all the stars in the $V$ vs ({\em B-V}) diagrams.
     {\bf Central panels}: all the stars in the $V$ vs ({\em V-I}) diagrams.
     {\bf Right panels} : Stars lying within $r$ arcmin from the cluster
     center (indicated for each cluster). The dashed line is the empirical
     ZAMS from Schmidt-Kaler (1982), whereas the solid lines are isochrones
     from Girardi et al.(2000) for the solar metallicity and the indicated age.
     }
     \end{figure*}

\section{Colour-Magnitude Diagrams and Cluster Parameters} 
In this section we describe cluster CMDs
and derive their basic parameters.

We first evaluated the CMD data as well as images from the 2-Micron All Sky 
Survey (2MASS) all sky data release (available at 
www.ipac.caltech.edu/2mass/releases/allsky) to explore the
existence of the clusters as physical systems.  The 2MASS $K_s$\ images are 
substantially less affected by reddening than the visual images which means 
that in some cases, the confusion from background galactic stars can be higher,
but the background should often also be less variable.

Distance moduli, reddenings  and ages of the confirmed clusters have been 
derived by matching the
observed CMDs to isochrones from the Padova group (Girardi et al. 2000)
by eye, paying
particular attention to the shape of the MS, the position of the brightest
MS stars, the turn-off point and the location of evolved stars, if
present.\\ 
To infer the heliocentric distances we adopted $R_V=A_V/E(B-V) = 3.1$

The theoretical isochrones are available for a wide range of metallicities.
We have adopted for all clusters those with a solar value because of a lack of
firm photometric or spectroscopic determinations of metallicities of
individual clusters. The
effect of metallicity has been frequently considered in literature:
increasing it shifts the isochrones fitting toward older ages, larger
distances and smaller reddening.

The results are summarized in Table~4, where the basic parameters
are listed together with their uncertainties. The latter correspond to the
shift allowed to isochrone fitting before a mismatch is clearly perceived
by eye inspection.\\

     \begin{figure*}
     \centerline{\psfig{file=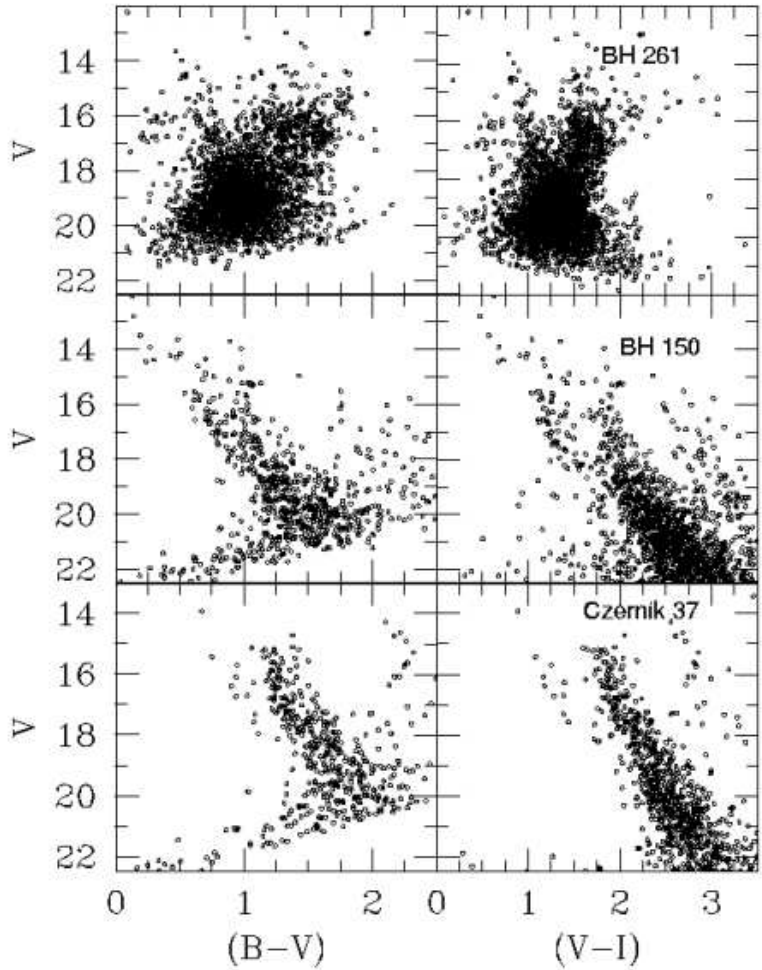,width=14cm}}
     \caption{
     CMDs of the stars in the field of Czernik~39, BH~150 and NGC~261. 
     {\bf Left panels}: all the stars in the $V$ vs ({\em B-V}) diagrams.
     {\bf Right panels}: all the stars in the $V$ vs ({\em V-I}) diagrams.
     }
     \end{figure*} 

\subsection{AL~1}
This cluster was discovered by Andrews \& Lindsay (1967) and then
independently by van den Bergh \& Hagen (1975) who named it BH~144.
It is described as a faint, moderately populated object, with a diamater 
of 1.5 arcmin,
clearly visible both on red and on blue plates (see Fig.~1).  The 2MASS $K_s$
image also shows the cluster just at the limiting magnitude of the image.
Its CMDs are shown in Fig.~9, lower panels.
Both the left-hand side and central panel clearly show the presence
of a strong contamination by the Galactic disk in a form resembling a
Main Sequence  (MS). The cluster lies on the left of the disk MS.
The evolved region is also complicated by foreground contamination.
However a selection in radius (0.6 arcmin) makes the cluster emerging.
A suggested match to an 800 Myr isochrone yields a reddening E(B-V)=0.35 and
a distance modulus (m-M) = 17.2. This implies a distance from the Sun
of 19.9 kpc. \\

\subsection{Berkeley~80}
This cluster was detected by 
Setteducati \& Weaver (1960). Dias et al. (2002) report
a diameter of 4 arcmin for Berkeley~80 (see also Fig.~2).  A small group of 
stars is visible at this position on the 2MASS $K_s$\ image.
Its CMDs are shown in Fig.~9, middle panels and resemble 
somewhat the CMD of NGC~5764 (see Fig.~9).
A possible match to a 300 Myr isochrone suggests a reddening E(B-V)=1.1 and
a distance modulus (m-M) = 14.8. This implies a distance from the Sun
of 3.3 kpc.\\

\subsection{NGC~5764}
This cluster is also listed as BH~167 by
van den Bergh \& Hagen (1975)
who described it as a very 
poorly populated object, with a diameter of 2.5 arcmin,
visible only on blue plates.  The cluster is also 
visible on our I-band image (see Fig 3) and the 2MASS $K_s$
image.  Its CMDs are shown in Fig.~9, upper panels.
Because of its distance from the Galactic Plane, this cluster
better emerges from the background, showing a clear MS from V=14 to V=
19.
The right-hand panel of Fig.~9 shows the central 1 arcmin region of the 
cluster region matched to a 200 Myr isochrone, 
yielding a reddening E(B-V)=1.0 and
a distance modulus (m-M) = 15.3. This implies a distance from the Sun
of 2.8 kpc

\subsection{Czernik~37}

     \begin{figure*}
     \centerline{\psfig{file=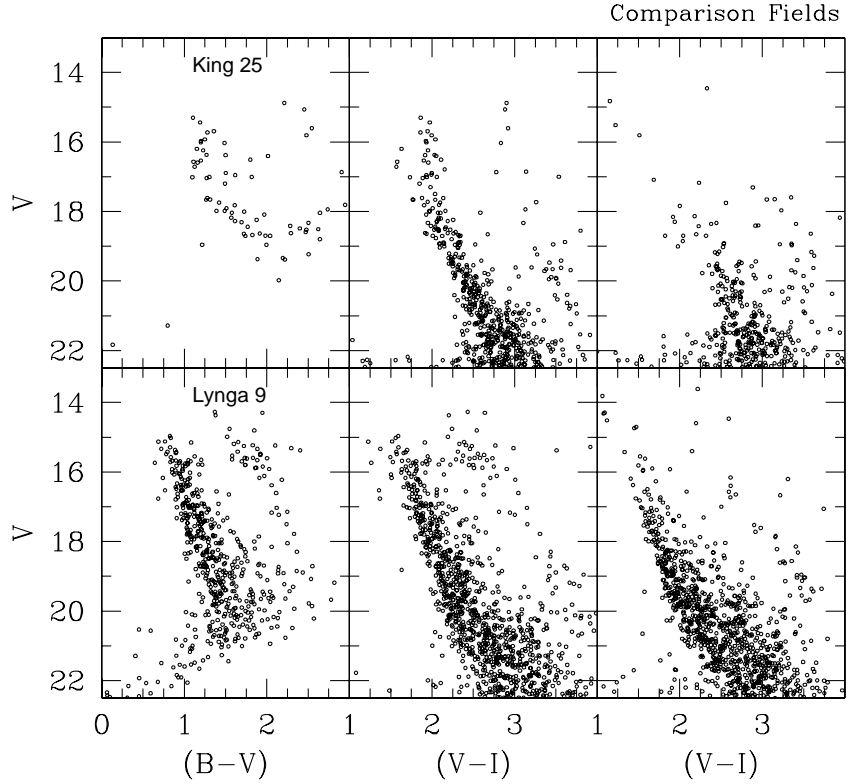,width=12cm}}
     \caption{  
     CMDs of the stars in the field of Lynga~9 and and King~25 and their
offset fields.
     {\bf Left panels}: all the cluster stars in the $V$ vs ({\em B-V}) 
diagram.
     {\bf Middle panels}: all the cluster stars in the $V$ vs ({\em V-I}) 
diagram.
     {\bf Right panels} : all the field stars in the $V$ vs ({\em V-I}) 
diagram
     }
     \end{figure*}

Czernik (1966) described this as a cluster with diameter of 3 arcmin
and fewer than 50 stars (see also Fig.~4).
The cluster is also listed as BH~253 by
van den Bergh \& Hagen (1975), who described it as a moderately populated 
object with the same diameter.  This cluster is projected onto the central
bulge of the galaxy, only 2 degrees from the direction to the galactic center.
The 2MASS image shows an extremely dense star field, with an 
asterism very much like the pattern visible in Fig.~4.  
Its CMDs are shown in Fig.~10, lower panels. Although there is an appearance 
of a broad main sequence in the CMD, it may just be an artifact of the
increasing star density in this direction.  The status of this possible cluster
is rather uncertain.
\\

\subsection{BH~150}
van den Bergh \& Hagen (1975) described this cluster 
as a faint, poorly populated object, with a diameter of 2.5 arcmin,
visible only on blue plates, but questionable in the red
plate.  In Fig.~5 and the 2MASS $K_s$\ image a bright star appears right 
at the cluster position, which may cause the impression of a small cluster.
Its CMDs are shown in Fig.~10, middle panels.  The CMDs for this region are 
dominated by the galactic plane population along this line of sight.
The evidence for a cluster at this position is weak at best.

\subsection{BH~261}
This possible cluster was discovered by 
van den Bergh \& Hagen (1975),
who describe it as a moderately populated cluster having
a diameter of 1.5 arcmin, clearly visible both on red and on blue
plates.  Fig.~6 shows no indication of a cluster, nor is there any 
sign of a cluster on the 2MASS $K_s$ image.
Its CMDs are shown in the upper  panels of Fig.~10.
The galactic contribution is very large and the field
very rich.   There probably is no cluster at this position.

\begin{figure} 
\centerline{\psfig{file=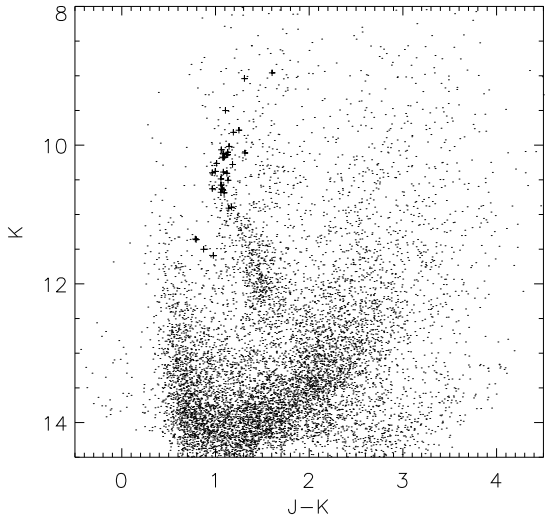,width=\columnwidth,angle=0}} 
\caption{K vs J-K cmd for 2mass stars within a 10 arcminute field of view
around the Lynga 9 position. The plus signs
indicate the Lynga 9 "clump" stars from Fig. 11 with
V between 14.5 - 16.5 and V-I between 2 - 3}
\end{figure} 

\subsection{Lynga 9}
This asterism was first noted by 
Lynga (1964), and later by van den Bergh \& Hagen (1975), who
named it BH~189, and
described it as a moderately populated cluster having
a diameter of 6 arcmin, clearly visible both on red and on blue
plates (see Fig.~7).  The 2MASS images show no indication of a cluster.
Its CMDs are shown in Fig.~11, together with a comparison 
field (upper right panel)
taken 10 arcmin away from the cluster center (see Sect.~2).  

Since we have the comparison field images for Lynga 9, it is possible to 
go into somewhat greater depth in our analysis of Lynga 9
than we can with the other clusters.
Assuming that the overall distribution of
stars in the field region is the same as the galactic background in the cluster
field, it is 
possible to subtract possible field stars from the cluster region with the 
following simple procedure.  For each 
star in the field region,
we simply deleted whatever star in the cluster region is closest 
in color and magnitude to the target field star.  
If the general background of stars is the same in the field region and the 
cluster region, and if there really is a cluster, this process would leave an 
excess of stars in the cluster field, most of which would presumably be
cluster stars.  

In the case of Lynga 9, the 
star-subtraction process eliminates all stars in the cluster region except 
for a few stars at the top of the ``main sequence'' and a group of stars that 
resemble a red giant clump at V=15.5 and
V-I=2.5.  Fainter than about V=16, the field region and the cluster region are
virtually identical.  
If the apparent ``clump'' stars really were red giants, there would have to be
a corresponding well-populated main sequence in the cluster region.
So there just cannot be an ordinary
cluster there.

So what is that clump of stars?  There certainly is
the appearance of a sparse cluster on the red and blue sky survey images, but
the 2MASS images show no indication of a cluster at all. 
However, a
K,J-K color-magnitude diagram for a field with a 10-arcminute radius around 
the cluster shows a prominent population of
stars near K=12 and J-K=1.5, together with a sequence of stars extending
brighter and to the blue (Fig. 12).

This large infrared clump so visible in Fig. 12 
is not the same as the Lynga 9 clump that appears in Fig. 11; the
Lynga 9 clump stars are all right at the top of the extension to the IR clump
near K=10, represented by + signs in Figure 12.  
Furthermore, the IR feature is distributed far more
widely than the cluster region
to at least a degree away in each direction from
the Lynga 9 position.

A possible explanation for Lynga 9 is that in this direction, near
the galactic plane, looking toward the galactic center, the star density
increases rapidly with distance.  So the apparent "main sequence" consists
of stars roughly at the far distance along the line of sight.  At some
distance along this line of sight, 
there may be a dense cloud, or possibly an entire spiral arm,
obscuring the view beyond. The "clump" stars of Figure 11 (as well as the
much larger group of stars that show up in the 2MASS data), may consist of
early-type stars behind, or more likely embedded in, this extended cloud
region.  What appears to be a cluster could be just a small window where the
obscuration is somewhat lower so that some of the brighter
stars in the cloud become visible.  
In all three of the Lynga 9 diagrams,
including the offset field, there is a sequence parallel to the apparent main
sequence - the colors of this sequence are consistent with being stars of the
same type as the "clump," but just more heavily reddened.

\subsection{King 25}
This cluster was detected by 
King (1966), who suggests it has a diameter of 5 arcmin,  
and is moderately populated (see Fig.~8).
There is no 
suggestion of a cluster on the 2MASS images.  
Its CMDs are shown in Fig.~11, 
together with a comparison field (upper right panel)
taken 10 arcmin apart from the cluster center (see Sect.~2).  Although the 
King 25 comparison field CMD appears different from that of the cluster field,
the cluster CMD strongly resenbles those of
several of the other clusters in the this sample.  The broad ``main sequence''
is likely to be entirely, or primarily the rich milky way background.  There 
probably is no cluster.

\begin{table*}
\caption{Parameters of the studied clusters. The coordinate system
is such that
the Y axis connects the Sun to the Galactic Center, while the X axis is 
positive in the direction of galactic rotation.
Y is positive toward the Galactic anti-center, 
and X is positive in the first and 
second Galactic quadrants (Lynga 1982).}
\fontsize{8} {10pt}\selectfont
\begin{tabular}{ccccccccc}
\hline
\multicolumn{1}{c} {$Name$} &
\multicolumn{1}{c} {$E(B-V)$}  &
\multicolumn{1}{c} {$(m-M)$} &
\multicolumn{1}{c} {$d_{\odot}$} &
\multicolumn{1}{c} {$X_{\odot}$} &
\multicolumn{1}{c} {$Y_{\odot}$} &
\multicolumn{1}{c} {$Z_{\odot}$} &
\multicolumn{1}{c} {$R_{GC}$} &
\multicolumn{1}{c} {$Age$} \\
\hline
& mag & mag& kpc & kpc & kpc & pc & kpc & Myr \\
\hline
AL~1        & 0.34$\pm$0.05  & 17.2$\pm$0.2 &16.9 & -13.8&  -9.8 & -950 & 
13.9&  800$\pm$200\\
NGC~5764    & 1.00$\pm$0.05  & 15.3$\pm$0.2 & 2.8 &  -1.7&  -2.2 &  290 &  
7.1&  200$\pm$100\\
Berkeley~80 & 1.10$\pm$0.05  & 14.8$\pm$0.2 & 3.3 &   1.7&  -2.8 &  -70 &  
7.3&  300$\pm$100\\
Czernik~37  & Possible cluster  &   &  &    &    &    &   &  \\
BH~150      & No cluster  &   &  &    &    &    &   &  \\
BH~261      & No cluster  &   &  &    &    &    &   &  \\
Lynga~9     & Spiral arm?  &   &  &    &    &    &   &  \\
King~25     & No cluster  &   &  &    &    &    &   &  \\
\hline
\end{tabular}
\end{table*}

\section{Discussions and Conclusions} 
The derived parameters of the program clusters are listed in 
Table~4.
Together with reddening, distance, age and the corresponding uncertainties,
we list the Galactocentric distance, derived by assuming $R_{\odot}=8.5 kpc$
and the Galactocentric rectangular coordinates $X_{\odot}$, $Y_{\odot}$ and 
$Z_{\odot}$.
The adopted reference system is centered on the Sun, 
with the X and Y axes lying on the Galactic plane
and Z perpendicular to the plane.
X points in the direction of the Galactic rotation, being positive in the first
and second Galactic quadrants; Y points toward the Galactic anticenter, 
being positive in
the second and third quadrant; finally, 
Z is positive towards the north Galactic pole (Lynga 1982).\\

AL~1 and NGC~5764 are located much further from the plane
than the thin disk mean scale-height; this is particularly interesting
for NGC~5764, whose age is unexpected for a cluster located 300
pc above the plane. AL~1 is a very interesting object, of intermediate age,
in the fourth quadrant, but very far from the galactic center and from
the galactic plane. 

Berkeley~80 see (Fig.~7)
and NGC~5764 (see Fig~3) have elongated shapes which might indicate they are
undergoing strong tidal interaction with Milky Way;

This work highlights the difficulties of working with open clusters towards
the inner regions of the galaxy.  The star densities are large, and increasing
rapidly with distance.  That causes the appearance of a main sequence on all
of the CMDs in this paper, resulting simply from the geometry of the situation.
Furthermore, patchy obscuration is likely to play a role in creating 
apparent ``clusters'' that are not physical associated groups of stars.  

Nevertheless, in recent work (Carraro et al 2005a,b,c) we have discovered
a considerable number of neglected intermediate-age open clusters,
which are going to significantly modify the open cluster age distribution
and probably the typical open cluster lifetime as presently known.

\section*{Acknowledgements} 
The observations presented in this paper have been carried out at
Cerro Tololo Interamerican Observatory CTIO (Chile).
CTIO is operated by the Association of Universities for Research in Astronomy,
Inc. (AURA), under a cooperative agreement with the National Science Foundation
as part of the National Optical Astronomy Observatory (NOAO).
The work of G. Carraro is supported by {\it Fundaci\'on Andes}.
This study made use of Simbad and WEBDA databases.  This publication makes use 
of data products from the 2-Micron All Sky Survey. which is a joint project of 
the University of Massachusetts and the Infrared Processing and Analysis 
Center/California Institute of Technology, funded by the National Aeronautics
and Space Administration and the National Science Foundation.
 


\begin{thebibliography}{} 
\bibitem{} Andrews A.D., Lindsay E.M. 1967, Irish Astron.
Journal 8, 126
\bibitem{} Baume G., Vazquez R.A., Carraro G., 2005, MNRAS 335, 475
\bibitem{} van den Bergh S., Hagen G.L. 1975, AJ 80, 11
\bibitem{} Bonatto, C., Bica, E., Pavani, D. B.  2004,
           A\&A 427, 485
\bibitem{} Carraro G., Girardi L., Marigo P., 2002, MNRAS 332, 705
\bibitem{} Carraro G., Geisler D., Baume G., Vazquez R.A., Moitinho A.,
           2005a, MNRAS 360, 655
\bibitem{} Carraro G., Mendez R.A., Costa E., 2005b, MNRAS 356, 647
\bibitem{} Carraro G., Baume G., Vazquez R.A., Moitinho A., Geisler D., 2005c,
           MNRAS, in press ({\tt astro-ph/0506694})
\bibitem{} Czernik M. 1966, Acta Astron. 16, 93
\bibitem{} Dias W.S., Alessi B.S., Moitinho A., Lepine J.R.D., 2002, 
           A\&AS 141, 371
\bibitem{} Girardi L., Bressan A., Bertelli G., Chiosi C., 2000, 
           A\&AS 141, 371
\bibitem{} Hasegawa T., et al., 2004, PASJ 56, 295
\bibitem{} Kaluzny J., 1994, A\&AS 108, 151 
\bibitem{} Kassis M., Janes K.A., Friel E.D., Phelps R.L., 1997, AJ 113, 1723
\bibitem{} King I., 1966, PASP 78, 81 
\bibitem{} Landolt A.U., 1992, AJ 104, 340 
\bibitem{} Lynga G., 1964, Lund Medd. Astron. Obs. Ser. II, 140, 1
\bibitem{} Lynga G., 1982, A\&A 109, 213
\bibitem{} Patat F., Carraro G., 2001, MNRAS 325, 1591 
\bibitem{} Phelps, R.L., Janes, K.A., Montgomery, K. A., 1994,
           AJ  107, 1079
\bibitem{} Schmidt-Kaler, Th. 1982, Landolt-B\"ornstein, Numerical data and Funct
    ional Relationships in Science and Technology, New Series, Group VI, Vol. 2(b),
    K. Schaifers and H.H. Voigt Eds., Springer Verlag, Berlin, p.14
\bibitem{} Setteducati A.E., Weaver M.F., 1960, in Newly found stellar clusters,
Radio Observatory Lab., Berkeley
\bibitem{} Stetson P.B., 1987, PASP 99, 191
\bibitem{} Wielen R., 1971, A\&A 13, 309
\end{thebibliography}
\end{document}